# Unifying Lexicons in view of a Phonological and Morphological Lexical DB

Monica Monachini[1]   Federico Calzolari[1,2]   Michele Mammini[1]

Sergio Rossi[1]   Marisa Ulivieri[1]

[1]Istituto di Linguistica Computazionale, CNR, Pisa
Via Moruzzi 1 – 56124 Pisa – Italy
e-mail: monica.monachini@ilc.cnr.it

[2]Scuola Normale Superiore, Pisa
Piazza dei Cavalieri 7 – 56126 Pisa – Italy
e-mail: federico.calzolari@sns.it

**Abstract**
The present work falls in the line of activities promoted by the European Languguage Resource Association (ELRA) Production Committee (PCom) and raises issues in methods, procedures and tools for the reusability, creation, and management of Language Resources. A two-fold purpose lies behind this experiment. The first aim is to investigate the feasibility, define methods and procedures for combining two Italian lexical resources that have incompatible formats and complementary information into a Unified Lexicon (UL). The adopted strategy and the procedures appointed are described together with the driving criterion of the merging task, where a balance between human and computational efforts is pursued. The coverage of the UL has been maximized, by making use of simple and fast matching procedures. The second aim is to exploit this newly obtained resource for implementing the phonological and morphological layers of the CLIPS lexical database. Implementing these new layers and linking them with the already exisitng syntactic and semantic layers is not a trivial task. The constraints imposed by the model, the impact at the architectural level and the solution adopted in order to make the whole database 'speak' efficiently are presented. Advantages vs. disadvantages are discussed.

## 1. Background and Motivations

The work described here raises issues in methods, procedures and tools for the reusability, creation, and management of Language Resources (LRs) and has been performed under the aegis of the European Language Resources Association (ELRA). ELRA is one of the major driving force and catalyst of a series of activities linked to LRs for the Human Language Technology sector. In the last years, one of its missions has been the production of LRs, fostered also via the packaging and customisation of already existing resources (www.elra.info). The present work falls in this line of activities promoted, specifically, by the ELRA Production Committee (PCom).
The idea behind the work is to conduct an experiment with a twofold purpose. From a purely methodological perspective, the aim is to investigate the feasibility, define methods and procedures for pooling and unifying two independently created Italian lexical resources into a Unified Lexicon (UL). This allows to combine two sources containing complementary information and incompatible formats at reasonable costs in terms of human efforts and computational techniques. From a concrete point of view, the experiment offers as a positive side-effect the possibility to exploit the new obtained resource for enriching an already existing lexicon with further linguistic modules. The implementation of a phonological and a morphological layer within the CLIPS architecture (Ruimy *et al.*, 2002; Ruimy *et al.*, 2003), a relational database which already contains the syntactic and semantic levels, is not a trivial task. The constraints imposed by the CLIPS model have interesting architectural impacts and force some implementation choices, in order to make the whole database 'speak' efficiently.
The paper mirrors this bipartition: two separate sections are dedicated, respectively, to the description of the Unified Lexicon experiment and the implementation of the phonological and morphological lexical layers.

## 2. The Unified Lexicon Experiment

Literature reports about some different approaches and methods for creating large-scale resources by combining already available sources. The trend is generally towards a semi-automatic approach. Chan and Wu (1999) present an intuitive and computationally not heavy method to merge lexicons that have incompatible Part-of-Speech categories: the merging is done via a set of mapping rules that compare each other the tags of those lemmas shared by both lexicons. Skoumalová (2001), embracing this trend, discusses the solutions adopted to combine existing resources with different formats to obtain an electronic syntactic lexicon of Czech: an automatic procedure of conversion from the source dictionary to the proposed format is appointed where, nevertheless, an amount of post-editing effort is required. Such exercises teach that the adopted methodologies and the consequent results strongly depend on (i) the aims of the task and (ii) the intrinsic characteristics of the resources involved. Moreover, the properties of the language are another noticeable variable which comes into play. All these factors constitute, hence, a barrier to the creation of a standard protocol for this by then consolidated orientation in resource building. As a consequence, everyone tends to develop the best suited strategies and ad-hoc procedures, where, generally, a balance between computational and human efforts is pursued.
This is also the driving criterion of the UL experiment proposed here, where incompatibilities and differences contained in the sources have been surmounted, trying to maximize the coverage of the UL with reasonable human efforts and not heavy nor sophisticated computational techniques, by a simple and fast matching procedure.

## 2.1 The Sources

The merging task has been carried out on two Italian lexicons available at ILC:
- a pronunciation lexicon, the DMI (Calzolari *et al.*, 1983), an Italian Machine Dictionary, containing, among other data, the phonological encoding;
- the Italian morphological module of the multi-layered lexicon PAROLE lexicon (Ruimy *et al.*, 1998).

The lexicons involved in this task are an ideal test for this kind of experiment: each source contains information not present in the other and, moreover, some data overlaps. Besides the complementary information, the two lexicons also present different formats. The DMI is substantially a list of inflected word-forms with information on:
- orthography of forms and lemmas (*pesca; pesche*)
- phonological encoding of word-forms and lemmas in a proprietary encoding:
    - vowel quality (1= closed, 2 = open vowel)
    - position of the accent (in the accent field, a number indicates the letter to be accented)
    - consonant quality (*razza*, voiced /unvoiced /*z*/)
- grammatical category of lemmas (e.g. N for Nouns)
- morphological features of inflected word-forms (e.g. FS for Feminine Singular)

PAROLE has a more sophisticated structure, being in SGML format[1]. It does not provides inflected word-forms but only lemmas, *Morphological_Units* (MUs), pointing to an inflectional pattern (*Ginp*), where inflectional rules generate all the inflections. The underlying philosophy relies on operations of the "*remove-add*" type.

The objective is that, at the end of the matching exercise, the PAROLE data, augmented with the pronunciation information from the DMI, will converge in the UL.

## 2.2 Balanced-Matching Procedure

Merging two lexicons with incompatible formats means to be confronted with two possibilities: (i) leave the native formats untouched, define specialized routines for getting the required data and moving them to the target source; (ii) choose one format as the leading one and conform the other to it, before merging them. Solution (ii) has been adopted where ASCII records has deemed the best work-format: PAROLE has been, hence, conformed to the DMI and the two files have been made as much parallel as possible, through different steps. First, an automatic procedure has been appointed to generate for the PAROLE lemmas the whole inflectional paradigms in the form of fixed-length field ASCII records, containing the *lemma, word-form, grammatical encoding and inflectional pattern*. A post-editing cycle has been carried out in order to deal with some inconsistencies: adjustments to flatten minor different orthography conventions (stress encoding), cancel mismatches in the αβ sorting of word-forms (the verbal paradigm, appearing in the αβ sorting of word-forms, in the one case, and in order of inflection, in the other case), normalize the number of word-forms belonging to a paradigm (e.g. the present participle word-forms). A further human intervention has been required to unify some small differences in content and format of the two PoS tagsets[2]. This cycle of not heavy even if time-consuming human intervention paved the way towards the next phase and the adoption of a rather simple automatic procedure.

At this point the two sources are ready to undergo a *balanced-matching* procedure. The matching conditions imposed are sufficiently restrictive, being the mapping calculated at the level of a window '*lemma-word-tag*'[3]. As a result, all the strings '*lemma-word-tag*' from PAROLE that match with a correspondent in DMI, converge in the new resource, where the inflectional encoding from the first, and the pronunciation information, from the latter are combined.

## 2.3 Italian Homographs but not Homophones

It should be noted that in PAROLE, homographic and not homophonous lemmas[4], e.g. *"peska* vs. *"pEska*, when belong to the same inflectional class, as the model imposes, receive a unique *Morphological_Unit*. Such a unit, once inflected, give rise, hence, to one set of word-forms only: therefore, the matching of PAROLE against the DMI, is able to retrieve phonological information relevant to one homograph only. The encoding of homographs (amounting to 648 total word-forms, out of 471544 matched) for which pronunciation information lack, is recovered by a manual post-editing.

## 2.4 The Unified Lexicon: Some Statistics

The results obtained by the adopted merging methodology are shown in Table 1:

| **DMI** | lemmas | 95119 |
|---|---|---|
| | word-forms | 1068937 |
| **PAROLE** | lemmas | 59881 |
| | word-forms | 515438 |
| **Unified Lexicon** | lemmas | 48208 |
| | word-forms | 472192 |

Table 1. UL coverage

The procedure appointed to match PAROLE against the DMI in view of their merging, is able to provide the phonological coverage to the 91.6% of the data[5]. Advantages and disadvantages, which are, obviously, inherent to each strategy, can be reported here as well. The generation of the morphological inflections and the conversion of PAROLE from SGML to the fixed-length field ASCII strings, which implies an automatic procedure to be appointed, can be seen as a drawback: however, this has a positive counterbalance, since it avoids writing more sophisticated routines to pick up information formatted in SGML. The post-editing phase necessary to reach the highest compatibility between the two sources, which is time-consuming in terms of human costs, can be also counted among inconveniences. Conversely, these efforts allow gaining in effectiveness and maximizing the coverage, by using simple computational techniques.

---

[1] The PAROLE morphology grounds on the GENELEX architecture (GENELEX 1993) and the EAGLES guidelines.

[2] It should be noted that the intervention on the tagsets has been very slight, since they are very compatible, the PAROLE tagset being EAGLES-conformant and the DMI tagset perfectly EAGLES mappable (Monachini and Calzolari 1996).

[3] The merging criterion adopted by Chan and Wu (1999) is based on the lemmas found in common in the two lexicons.

[4] These cases will turn out of interest in the second phase of the work, where they will influence the architectural choices in the implementation of the morphological layer.

[5] The 8.4% of data can remain unmatched because of two reasons: (i) the PAROLE lemma does not exist in the DMI or (ii) the windows do not match.

## 3. The CLIPS Phonological and Morphological Database

This section deals with the architectural and implementation solutions adopted for augmenting the CLIPS database with two new linguistic layers.

### 3.1 The Model

The Unified Lexicon offers the possibility to import two new modules in the CLIPS relational database, the phonological repository of word-forms with the pronunciation encoding and the morphological component with lemmas and their inflectional codes.

The model foresees four independent modules, all building on an *entity-relationship* philosophy (Fig.1). The entities of the phonological and morphological layers, the same as those of the semantic and syntactic ones, stand in *n:m* relation. However, the morphological entities stand in *1:n* relation with those of the syntactic layer. That being the way, the morphology constitutes a *bottleneck* that blocks the flow of information between the syntactic and/or semantic levels and the phonological one. Specifically, homographic but not homophonous entries can not access their correct phonological apparatus: let consider, indeed, that, both *pesca*[Fruit] and *pesca*[Sport], via the syntax, point to the morphological unit *pesca*_N and, through it, have access to "*peska* and "*pEska*, indistinctly.

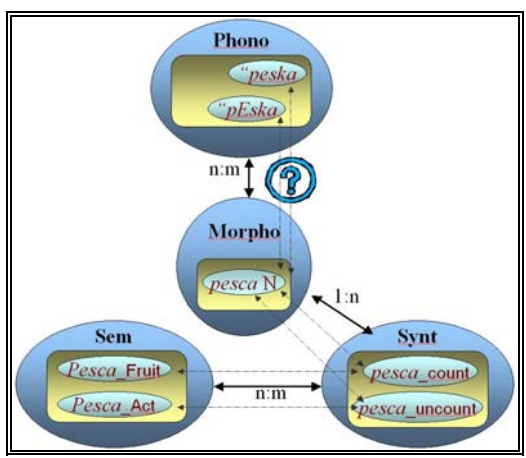

Figure 1. The model

### 3.2 The Architecture of the Database

The overall database architecture requires that the *n:m* relations between the units of the different layers are made explicit in a relevant correspondence table. The two incoming modules, first, should be made 'speak' one each other according to this architecture, by translating their *n:m* relations into such a table[6]. Then, the flow of information between the two new and the already existing layers should be made transparent and the bottleneck created by the morphology solved. Two possible architectural solutions can be envisaged. One possibility should be skipping the morphological level and defining correspondences between the semantic/syntactic units and the relevant phonological units, directly. This hypothesis contains, however, a disadvantage: for each semantic unit, relations should be stated with 'all' possible inflections, thus making the correspondence table heavy in terms of data redundancy and size, with negative repercussions on the database management. The second strategy foresees to violate, within the morphological component, the rule of non-redundancy (intrinsic to the entity-relationship model) in the creation of the morphological entries and to duplicate those homographic units (180 cases like *pesca*), pointed by non homophonous semantic units. This expedient creates a light data redundancy within the morphological layer[7], but, allows to use it, for the sake of speed and size[8], as a bridge between the new and already existing lexical layers.

### 3.3 The Implementation

The archives of both the morphological and phonological layers are stored starting from the UL. This lexicon allows to implement what can be defined the 'core' of the two layers, the correspondence table that relates each other their entities, i.e. the *Morphological_Units* (MUs) and *Phonological_Units* (PhUs). The tables of the PhUs and MUs are obtained accordingly, importing for each of them the relevant linguistic information: pronunciation encoding, for the first, and inflectional code (Ginp), for the latter.

In the phonological layer, the PHUs present another link as well, the connection to the *Phonological_Variants* (PhUVs), if any[9]. The phonological transcription can be shown either in the DMI proprietary format or in the into the computer-readable phonetic alphabet SAMPA (www.phon.ucl.ac.uk). Ad-hoc routines are developed to convert the two pronunciations each other.

In the morphological layer, the MUs point to the Ginp and, through it, to a set of *remove-add* rules implemented to generate the whole inflectional paradigms.

The bridge between the morphology and the syntax and, through it, to the subsequent level, the *Semantic_Units* (*SemUs*), is provided by means of a correspondence table, where the MUs are linked to the *Syntactic_Units* (SynUs). Figure 2. displays all such a links.

---

[6] The relations between the syntactic and semantic entries have been already implemented accordingly.

[7] The MU *pesca1* and *pesca2* share the same *Ginp* and there is no point to split them, except for phonological considerations (which 'corrupt' the pure morphological criterion).

[8] The increase in records with the first solution is calculated around the 40 pc.

[9] Out of 367991 different PhUs and 49071 MUs, the correspondence table PhUs-MUs amounts to 472017 records. For 2464 PhUs a transcription variant (PhUVs) is provided.

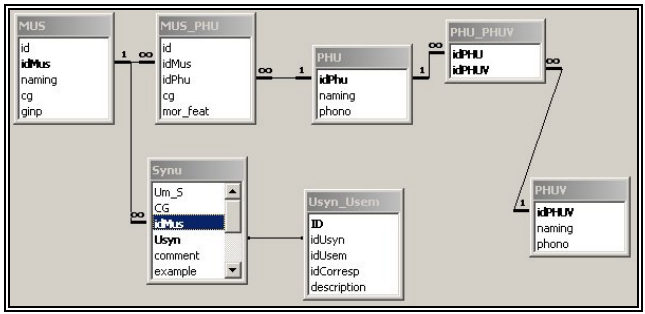

Figure 2. The Database

All the lexical entries, throughout the four levels, can be browsed independently by means of user-friendly form interfaces, implemented in the CLIPS browsing tool: being all the lexical layers interconnected, once entered an orthographic form (either a word-form or a lemma), it is possible to access all the linguistic information linked to it.

### 3.4 The Exportation

In order to make the two new layers exportable in XML-format, according to the whole database, the DTD for the phonological layer is created[10] and appropriate export procedures designed. An XML phonological entry contains the transcription information (both proprietary encoding and SAMPA alphabet), the correspondence with the possible morphological unit(s), the grammatical encoding, and, finally, the correspondence to the variants, if any. Via the morphological unit(s), it is connected to the subsequent linguistic layers, what allows a whole XML-entry to be obtained, from the phonological to the semantic levels (Fig. 3).

```xml
-<Parole lexiconname="CLIPS" language="Italian" integrity="WITHOUTB">
  -<ClipsPhono>
    -<PhU id="PHUpesca" naming="pesca" dmi="pe1sca" sampa=""peska">
       <CorrespMuPhu targetMu="MUSpescaNOUN2" gramcat="N" morph_feat="FS" />
     </PhU>
   </ClipsPhono>
  -<ParoleMorpho>
    -<MuS id="MUSpescaNOUN2" naming="pesca" gramcat="NOUN"
       synulist="SYNUpescaN2 SYNUpescaN3" gramsubcat="WITHOUTSC"
       autonomy="WITHOUTB" >
       <GInP id="110" />
     </MuS>
   </ParoleMorpho>
  -<ParoleSyntaxe>
    -<SynU id="SYNUpescaN2" example="sorta di lotteria" naming="pesca">
       <CorrespSynUSemU targetsemu="USem60483pesca" />
     </SynU>
    -<SynU id="SYNUpescaN3" example="la pesca della balena da parte dei pescatori"
       naming="pesca">
       <CorrespSynUSemU targetsemu="USem60481pesca" />
     </SynU>
   </ParoleSyntaxe>
  -<ParoleSemant>
      <SemU id="USem60481pesca" naming="pesca" example="la pesca della balena da
       parte dei pescatori" freedefinition="il pescare" weightvalsemfeaturel="Act" />
      <SemU id="USem60483pesca" naming="pesca" example="pesca di beneficenza"
       freedefinition="sorta di lotteria a estrazione" weightvalsemfeaturel="Purpose_Act" />
   </ParoleSemant>
 </Parole>
```

Figure 3. An XML lexical entry

In the ILC Web site, a portion of the CLIPS syntactic and semantic data can be output in HTML-format. These are pages where information is stored statically, by exporting the data directly from the database. A future perspective is to provide with the possibility of linking either the phonology or the morphology and display the information in the same format, in order to allow an easier and faster human readability and navigability of the CLIPS data.

### 5. Concluding Remarks and Future Works

In the paper, the best strategy to combine two independently created lexical resources of Italian, in view of obtaining an Unified Lexicon is described. Obviuosly, the procedure should not be seen as a proposal for a standard protocol in such kind of activities: the method strongly depends on the sources coming into play and only is applicable to them. From the Unified Lexicon, two further linguistic layers of the CLIPS lexical database, phonology and morphology, are implemented. The overall model, the architecture of the whole relational database and the solutions at implementation level are presented.
In the future, a possible development of this work would be to make the whole database interactively consultable on internet. In the web application, the dynamic navigation of the data throughout the linguistic layers could be allowed and web pages dynamically generated. In this perspective, the possibility (or better the necessity) to make use of a database which optimizes the web performances and increases the speed of access should be object of a careful evaluation.

---

[10] The DTDs for the morphological, syntactic and semantic levels were already available from PAROLE.